\documentclass[twocolumn,showpacs,preprintnumbers,amsmath,amssymb]{revtex4}

\usepackage{graphicx}
\usepackage{dcolumn}
\usepackage{bm}

\begin{document}


\title{Ultra-broadband photon pair preparation by spontaneous four wave mixing in dispersion-engineered optical fiber}

\author{Karina Garay-Palmett$^{1}$, Alfred B. U'Ren$^{1,2}$, Ra\'ul Rangel-Rojo$^{1}$,\\Rodger Evans$^{1}$ and Santiago Camacho-L\'opez$^{1}$}
\affiliation{
Departamento de Optica, Centro de Investigaci\'on Cient\'{\i}fica y
de Educaci\'on Superior de Ensenada, Apartado Postal 2732, Ensenada,
BC 22860, Mexico.\\
$^2$Instituto de Ciencias Nucleares, Universidad Nacional Autonoma de Mexico, apdo. postal 70-543, Mexico 04510 DF
}

\date{\today}

%
\newcommand{\epsfg}[2]{\centerline{\scalebox{#2}{\epsfbox{#1}}}}

\begin{abstract}
We present a study of the spectral properties of photon pairs
generated through the process of spontaneous four wave mixing (SFWM)
in single mode fiber.  Our analysis assumes narrowband pumps, which
are allowed to be frequency-degenerate or non-degenerate.  Based on
this analysis, we derive conditions on the pump frequencies and on
the fiber dispersion parameters which guarantee the generation of
ultra-broadband photon pairs.   Such photon pairs
are characterized by:  i) a very large degree of entanglement, and ii) a very high degree of temporal synchronization between the signal and idler photons. Through a numerical exercise, we find that the use of
photonic crystal fiber (PCF) facilitates the fulfilment of the
conditions for ultra-broadband photon pair generation;  in particular, the spectral region in which emission occurs can be adjusted
to particular needs through an appropriate choice of the PCF
parameters. In addition, we present a novel quantum interference effect, resulting from indistinguishable pathways to the same outcome, which can occur when pumping a SFWM source with multiple spectral lines.
\end{abstract}

\maketitle

\section{Introduction}

Quantum-enhanced technologies require photon pairs with specific
properties. Thus, for example, factorability is required for the
generation of pure heralded single photons, a crucial resource for linear optical quantum
computation\cite{uren05}. At the opposite extreme, spontaneous
parametric processes permit the generation of photon pairs with a
remarkably large degree of entanglement. Such photon pairs are
characterized by a large Schmidt number\cite{cklaw00}, which implies
that emission takes place into a large number of independent
frequency or transverse momentum signal and idler mode pairs. A
state with these characteristics leads to a large mutual
information, which quantifies the information which two parties can
in principle share by virtue of the entanglement
present\cite{zhang07}. Two-photon states with a large degree of
entanglement may lead to exciting applications, such as
large-alphabet quantum key distribution\cite{aKhan07},
quantum-enhanced two photon absorption\cite{Dayan04},  and
teleportation of single-photon wavepackets\cite{molotkov}.

Most research on photon pair generation has relied on the process of spontaneous parametric downconversion (PDC) in second-order non-linear crystals\cite{burnham70}.   Recently, the process of spontaneous four-wave-mixing (SFWM) in optical fiber, relying on a third-order non-linearity, has emerged as a useful alternative, with some clear advantages.  Indeed, fibers permit an essentially unlimited interaction length; in addition, the generated flux is proportional to the square of the pump power, instead of the linear pump-power dependence observed for PDC.   This leads to the possibility of remarkably bright sources. While some of the first SFWM experiments were carried out with standard telecom fiber\cite{wang01}, recent experiments have exploited the greater flexibility conferred by photonic crystal fibers (PCF) \cite{sharping04}.  Thus, for example, the zero group velocity dispersion frequencies may be selected by careful choice of the fiber design parameters, leading to the ability to also select the generation frequencies.  In previous work, we have shown\cite{garay07} that SFWM in PCF's leads to the ability to engineer the spectral entanglement properties of photon pairs.

We are particulary interested in spontaneous parametric processes
constrained to a single transverse mode, e.g. in a single-mode
waveguide or fiber.  Under these circumstances transverse momentum
entanglement as well as mixed spectral-transverse momentum
entanglement are automatically suppressed. Thus, for
single-transverse-mode parametric processes we may limit our
analysis of the entanglement present to the  spectral degree of
freedom. It has been shown that a large Schmidt number can then be
obtained by engineering the photon pair generation process to yield
the largest possible emission bandwidth  compatible with the smallest possible
pump bandwidth\cite{zhang07}.  Apart from its role in
enhancing the attainable degree of entanglement, a large generation
bandwidth also leads to a small correlation time, defined as the
width of the time of emission difference (between  signal and idler)
probability distribution\cite{Harris07,Strekalov05}.  A source with
these characteristics would be useful for applications relying on a
short time of arrival difference between two optical modes, such as
quantum optical coherence tomography\cite{nasr03} where the
instrument resolution is inversely proportional to  the correlation
time. Let us note that an ultra-broadband photon pair source treated
classically, i.e. where we employ standard non-photon-counting
detectors, could serve as a substitute for white light sources (e.g.
based on self-phase modulation in fibers) with one clear advantage:
while white light spectra often have a rather irregular shape,
ultra-broadband parametric processes can be engineered to have a
nearly flat spectrum.  Likewise, a source with these characteristics
could lead to parametric amplifiers with an exceptionally large
bandwidth.

The generation of ultra-broadband two-photon states  is possible
with sources based on parametric downconversion (PDC) relying on
second-order non-linear crystals\cite{Nasr05,Carrasco06,nasr08,hendrych08}.   In a
recent experiment, it was shown that selecting the non-linear
crystal and pump frequency so that the emitted light is centered at
the zero group velocity dispersion frequency of the non-linear
medium leads to PDC light with a remarkably broad spectrum; a full
width at half maximum generation bandwidth of $1.080\mu$m centered
at $1.885\mu$m was demonstrated\cite{odonnell07}.  Nevertheless, an
important limitation of an approach based on PDC, is that because broadband emission occurs at the zero dispersion frequency, the central emission frequency cannot be freely selected.
Note that while periodic poling in non-linear crystals\cite{fejer92} can be exploited to yield phase-matching at arbitrary wavelengths, it cannot be used to manipulate the zero dispersion frequency, and hence to select the ultra-broadband central frequency; thus, the experiment discussed above operates in a spectral region which is inconvenient for many applications.  This
leads to the motivation for the present work, in which we aim to
develop fiber-based ultra-broadband photon pair sources, with
far greater freedom for selecting the central generation
frequency and other emission characteristics.

In this paper we analyze the spectral properties of photon pairs
generated through the process of spontaneous four wave mixing (SFWM)
in optical fiber.  We concentrate on the case of quasi-monochromatic
pumps which may be either frequency degenerate or non-degenerate,
and where all the fields propagate in the fundamental fiber mode.  While our theory can be applied to any  fiber, we illustrate our discussion with a numerical exercise for the specific case of photonic crystal fiber, composed of a solid core and a cladding with an array of air
holes. The main motivation for using PCF is that the resulting
dispersion properties can be tailored by careful choice of the fiber
design parameters including the size, location and shape of the air
holes, which in turn permits the generation of photon pairs with
tailored properties.  We analyze the conditions which must be imposed on the pump
frequencies and the fiber parameters to permit ultra-broadband
photon pair generation.    Note that while broadband four-wave-mixing in optical fiber has been analyzed from a classical perspective by several groups\cite{marhic96}, such an analysis has not been presented for the spontaneous, non-classical case.  We also discuss a quantum interference effect which can occur in the
process of SFWM when the pump includes multiple spectral lines, for
example corresponding to degenerate and non-degenerate pumps.

\section{Spontaneous four wave mixing theory}

We study the spontaneous four wave mixing process in single-mode
fiber with a third-order nonlinearity  $\chi^{(3)}$. In this
process, a photon pair, comprised of one photon in the signal mode,
$\hat{E}_{s}$, and one photon in the idler mode $\hat{E}_i$, is
created by joint annihilation of two photons from the pump fields
$E_{1}$ and $E_{2}$.  In this paper we focus on source geometries
which permit the generation of particularly broadband signal and
idler photon-pair states. Remarkably, this is
possible even if the pumps are nearly monochromatic.   Our theory in this paper is valid in the limit of monochromatic pumps, and for all generation frequencies.  In contrast, the theory presented in a related paper from our group\cite{garay07} for broadband pumps, is valid only for a narrow signal and idler spectral vicinity and is therefore unsuitable for the description of broadband photon-pair generation.   In our present analysis we assume that all fields are co-polarized and that they propagate in the fundamental transverse mode of the fiber.

The quantum state of the generated photon pair in an optical fiber
of length $L$ can be obtained following a standard perturbative
approach \cite{mandel} and is given by

\begin{align}
\label{state} &  |\Psi\rangle= |0\rangle_{s}|0\rangle_{i}
+\kappa\int\int d\omega_{s} d\omega_{i}F\left(
\omega_{s},\omega_{i}\right)  \left| \omega_{s}\right\rangle _{s}
\left|  \omega_{i}\right\rangle _{i},
\end{align}

\noindent where $\kappa$ is a constant which represents the
generation efficiency and $F\left(\omega_s,\omega_i\right)$ is the
joint spectral amplitude function (JSA), which describes the
spectral entanglement properties of the photon pair

\begin{align}
\label{eq: JSA}&F\left(  \omega_{s},\omega_{i}\right)     = \int
d\omega^{\prime}\alpha_{1}\left(  \omega^{\prime}\right)  \alpha_{2}
\left(
\omega_{s}+\omega_{i}-\omega^{\prime}\right) \nonumber\\
&  \times\mbox{sinc}\left[L\Delta k\left(  \omega^{\prime}%
,\omega_{s},\omega_{i}\right) /2 \right]  \mbox{exp} \left[ iL\Delta k\left(
\omega^{\prime},\omega_{s},\omega_{i}\right)/2 \right] .
\end{align}

The JSA function is given in terms of the phasemismatch function

\begin{align}
\label{deltak} \Delta k\left( \omega _{1},\omega _{s},\omega
_{i}\right) &=k\left( \omega _{1}\right) +k\left( \omega _{s}+\omega
_{i}-\omega _{1}\right) -k( \omega _{s}) \nonumber \\&- k( \omega
_{i}) -( \gamma _{1}P_{1}+\gamma _{2}P_{2}),
\end{align}

\noindent which includes self/cross-phase modulation contributions
for the two pumps with peak powers $P_{1}$  and $P_{2}$,
characterized by the nonlinear parameters $\gamma_1$ and $\gamma_2$;
$\alpha _i(\omega)$ (with $i=1,2$) represents the spectral shape of
the pumps. The energy conservation constraint is apparent in the
argument of the second term of the phase mismatch\cite{garay07}.

\subsection{Spontaneous four-wave mixing with narrowband pumps}

An expression for the JSA (Eq.(~\ref{eq: JSA})) in closed analytic
form, valid for all generation frequencies, can be derived if both pumps are nearly monochromatic.  Here we assume that the pumps have a
Gaussian spectral profile, centered at frequencies $\omega_{1}$ and
$\omega_{2}$, each with a narrow bandwidth $\sigma$. It can be shown
that in the limit where both pumps are monochromatic, i.e. $\sigma
\rightarrow 0$, the product of the two pump envelope functions in
Eq.(\ref{eq: JSA}) reduces to

\begin{align}
\label{deltas}&\alpha_1\left(\omega'\right) \alpha_2
\left(\omega_s+\omega_i-\omega'\right)\rightarrow
\pi\sigma^2\delta(\omega_s+\omega_i-\omega_{1}-\omega_{2})
\nonumber\\& \times
\delta\left[\omega'-\left(\omega_s+\omega_i+\omega_{1}-\omega_{2}\right)/2
\right].
\end{align}

The appearance of a Dirac delta function involving $\omega'$ in
Eq.(\ref{deltas}) allows us to carry out the integral in
Eq.(\ref{eq: JSA}). Thus, in the limit of monochromatic pumps, the JSA is given by

\begin{align}
\label{JSA1} &F_{cw}(\omega_s,\omega_i)=N
\delta(\omega_s+\omega_i-\omega_{1}-\omega_{2})\nonumber \\ &\times
\mbox{sinc}\left[L\Delta k_{cw}(\omega_s,\omega_i)/2 \right]
\exp\left[iL\Delta k_{cw}(\omega_s,\omega_i)/2 \right],
\end{align}

\noindent in terms of a normalization constant $N$ and the phase mismatch $\Delta
k_{cw}(\omega_s,\omega_i)$ which is now a function only of
$\omega_s$ and $\omega_i$

\begin{align}
\label{Dk} \Delta
k_{cw}(\omega_s,\omega_i)&=k\left[\left(\omega_s+\omega_i+\omega_{1}
-\omega_{2}\right)/2\right]\nonumber\\&+k\left[\left(\omega_s+\omega_i-\omega_{1}+
\omega_{2}\right)/2\right]\nonumber\\&-k(\omega_s)-k(\omega_i)-(\gamma_1
P_1+\gamma_2 P_2).
\end{align}

Making use of Eqns.~\ref{state}, \ref{JSA1}-\ref{Dk}, we can write
down the two photon state as

\begin{equation}
\label{eq: state}  |\Psi\rangle= |0\rangle_{s}|0\rangle_{i}
+\kappa\int d\omega f\left(
\omega\right) | \omega\rangle _{s}
 | \omega_{1}+\omega_{2}-\omega\rangle _{i},
\end{equation}

\noindent where

\begin{equation}
\label{espindiv} f(\omega)=N\mbox{sinc}[L \Delta
k_{cw,sing}(\omega)/2]\exp[iL \Delta k_{cw,sing}(\omega)/2],
\end{equation}

\noindent in terms of

\begin{eqnarray}
\label{delksing} &\Delta k_{cw,sing}(\omega)=\Delta
k_{cw}(\omega,\omega_{1}+\omega_{2}-\omega)=k(\omega_{1}) \nonumber
\\
&+k(\omega_{2})-k(\omega)-k(\omega_{1}+\omega_{2}-\omega)-(\gamma_1
P_1+\gamma_2 P_2).
\end{eqnarray}

Note that while $|F_{cw}(\omega_s,\omega_i)|^2$ represents the joint
spectrum, $|f(\omega)|^2$ represents the singles spectrum. From
Eq.(\ref{JSA1}), it is clear that photon-pair generation requires
the fulfilment of the following two conditions: i) energy
conservation, or $\omega_s+\omega_i=\omega_{1}+\omega_{2}$  and ii)
momentum conservation or $\Delta k_{cw} \approx 0$, with a tolerance
which is inversely proportional to $L$. In order to analyze the
phase matching properties near the zero group velocity dispersion
frequency $\omega_{zd}$ of the fiber, we express the propagation
constant as a Taylor series centered at
$\omega_s=\omega_i=\omega_{zd}$. Thus, the phasemismatch may be
expressed to fourth order as

\begin{align}\label{deltakcw}
&\Delta k_{cw}^{(4)} (\delta_{+},\delta_{-})= -(\gamma_1
P_1+\gamma_2 P_2) +\delta k_0+\frac{1}{2}\delta k_1\delta_+
\nonumber \\& +\frac{1}{4(2!)}\delta
k_2\delta_{+}^2+\frac{1}{8(3!)}\left[ \delta
k_3\delta_{+}^3-6k^{(3)}\delta_{+}\delta_{-}^2\right]\nonumber
\\
& +\frac{1}{16(4!)}\left[ \delta
k_4\delta_{+}^4-2k^{(4)}\left(6\delta_{+}^2-\delta_{-}^2\right)\delta_{-}^2\right],
\end{align}

\noindent where $\delta k_n = k^{(n)}_{+}+k^{(n)}_{-}-2k^{(n)}$,
where $k^{(n)}_\pm$ represents the $n$th frequency derivative of $k$
evaluated at
$\Omega_{\pm}=\omega_{zd}\pm(\omega_{1}-\omega_{2})/2$, and where
$k^{(n)}$ represents the $n$th frequency derivative of $k$ evaluated
at $\omega_{zd}$. Here, we have defined detunings with respect to
$\omega_{zd}$, and furthermore have defined new variables given by
the sum and difference of these detunings

\begin{eqnarray}
\label{detu}
\delta_{\mu}&=&\omega_{\mu}-\omega_{zd},\ \ \ \mu=s,i,1,2 \nonumber \\
\delta_\pm&=&\delta_s \pm \delta_i .
\end{eqnarray}

Let us note that the constant term of the expansion, $\delta k_0$,
vanishes if perfect phasematching is achieved for pump frequencies
$\omega_{1}$ and $\omega_{2}$ at the generated frequencies
$\omega_s=\omega_{zd}$ and $\omega_i=\omega_{zd}$. The JSA
can then be written as

\begin{eqnarray}
\label{jsadpmas} F_{cw}(\delta_{+},\delta_{-})&=&N
\delta(\delta_{+}-\delta_{p+}) \mbox{sinc}\left[L \Delta
k_{cw}^{(4)}(\delta_{+},\delta_{-})/2\right] \nonumber \\ && \times
\mbox{exp}\left[i L \Delta
k_{cw}^{(4)}(\delta_{+},\delta_{-})/2\right],
\end{eqnarray}

\noindent where we have defined $\delta_{p+}=\delta_{1}+\delta_{2}$.

\subsection{Step index dispersion model for photonic crystal fiber}

Our analysis so far, and the conditions for ultrabroadband photon pair generation to be presented in Sec.~\ref{condiciones},
are valid for any fiber.  In order to carry out specific numerical calculations, in this paper we use specialize our discussion to photonic crystal fiber.  This type of fiber consists of a fused silica core
surrounded by silica cladding with a pattern of air holes which
remains constant along the fiber length. This mixture of air and and
glass in the cladding results in an average refractive index that is
considerably lower than that of the core, providing a high
dielectric contrast, resulting in strong optical confinement. This
leads to high peak irradiances even for modest input powers, which
enhances nonlinear optical effects such as SFWM. In addition, the
dispersion characteristics of the PCFs can be engineered by
variations of the distribution, size and shape of the air holes
surrounding the core. In particular, it becomes possible to choose
the zero dispersion frequencies, to tailor the SFWM
phase-matching properties \cite{hansen03} and to design fibers
approaching endlessly single-mode behavior \cite{birks97}.

The SFWM phase-matching properties are determined by the fundamental
mode propagation constant, given in terms of the effective
refractive index $n_{eff}$ by $k(\omega )=n_{eff}(\omega )\omega
/c$.  We adopt a step-index model, where the core has radius $r$,
its index is that of fused silica $ n_{s}(\omega )$, and the
cladding index is calculated as $ n_{clad}(\omega
)=f+(1-f)n_{s}(\omega )$, where $f$ is the air-filling fraction.
According to Ref.~\cite{Wong2005} this fiber dispersion model is
accurate for $0.1 \le f \le 0.9$. In the context of our work, this
model permits a straightforward exploration of the spectral
entanglement properties in $\{r,f\}$ parameter space.

\subsection{Phase matching properties for degenerate pumps}\label{degpum}

Let us now restrict our treatment to the degenerate pumps regime,
i.e. $\omega_{1}=\omega_{2}=\omega_p$. In this case,
$\Omega_+=\Omega_-=\omega_{zd}$ and therefore, it can be inferred
that, except for the self/cross-phase modulation term, the series in
Eq.(\ref{deltakcw}) does not start until third-order terms and all
terms $\delta k_n$ vanish. Thus, with $P_1=P_2=P$ and $\gamma_1=\gamma_2=\gamma$,
Eq.(\ref{deltakcw}) reduces to

\begin{align}\label{dkdeg}
\Delta k_{cw}^{(4)} (\delta_{+},\delta_{-})&=-2\gamma
P-[\delta_{-}]^2\nonumber\\&\times\left[\frac{3 k^{(3)}}{4 (3!)}
\delta_{+} +\frac{k^{(4)}}{8 (4!)}(6
\delta_{+}^2+\delta_{-}^2)\right].
\end{align}

For a negligible self/cross-phase modulation contribution, an
analysis of Eq.(\ref{dkdeg}) reveals the existence of two distinct
phasematching ``branches'', where each one corresponds to one of the
factors in square brackets vanishing. We refer to the first as
the trivial branch; indeed together with energy conservation it may
be seen to be centered at $\omega_s=\omega_{i}=\omega_p$.

\begin{figure*}[t]
\begin{center}
\centering\includegraphics[width=14cm]{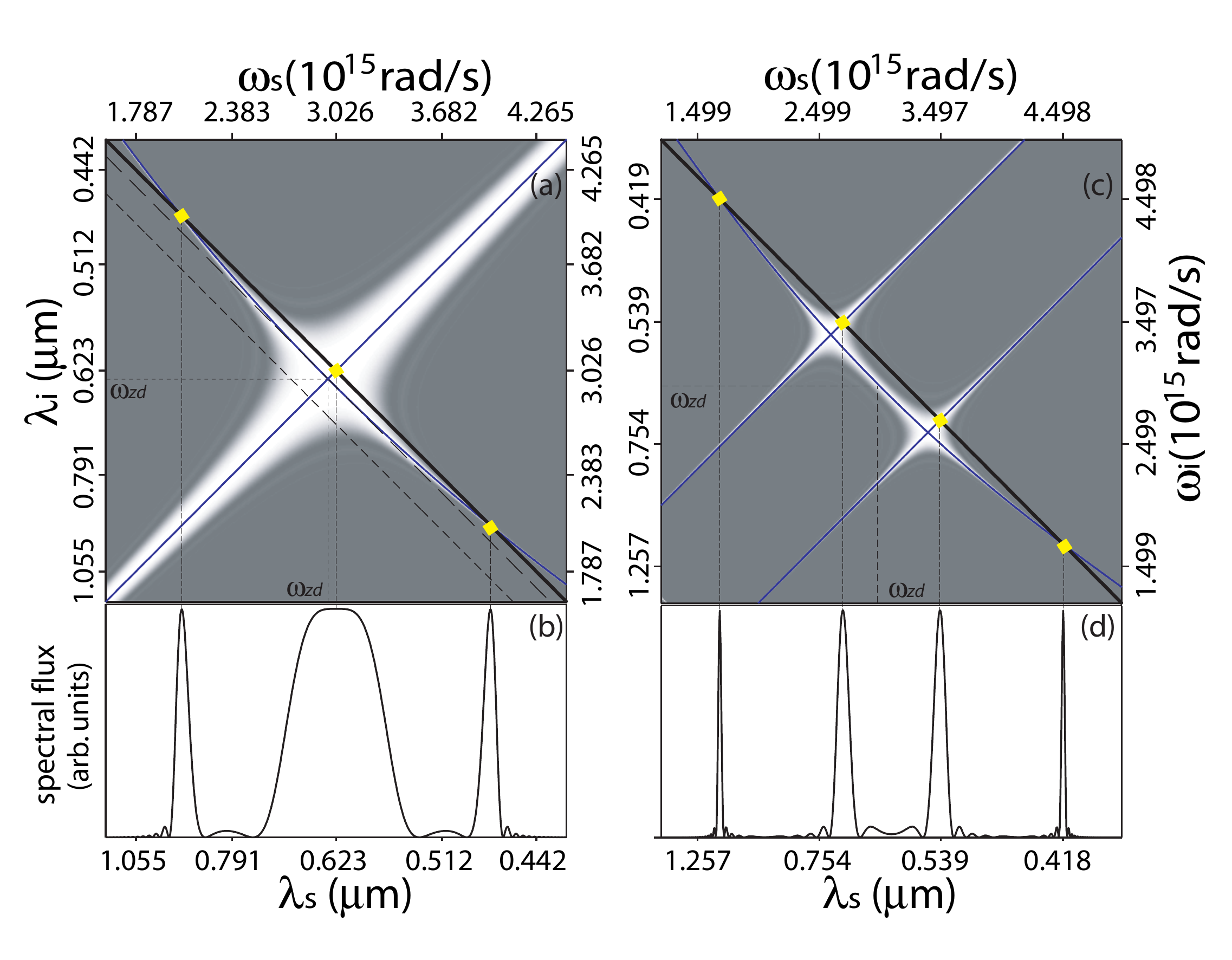}
\end{center}
\par
\caption{(Color online) Representation of the joint and singles spectrum
for a photonic crystal fiber with $r=0.7000 \mu$m and $f=0.9$,
leading to $\lambda_{zd}=2 \pi c/\omega_{zd}=0.6335\mu$m, in the low pump peak power
limit and with a fiber length of $L=1$cm. Note that frequency axes are
labeled for convenience with wavelength values. (a) Phasematching
function $|\mbox{sinc}[L \Delta k_{cw}(\omega_s,\omega_i)]/2|^2$ plotted as
a function of $\omega_s$ and $\omega_i$; lines with negative unit
slope represent energy conservation loci for different values
of $\omega_p$ around $\omega_{zd}$.  Thin blue lines represent the
perfect phasematching contour.  The joint spectrum is obtained as the
intersection of the energy
conservation locus with the phasematched region in $\{\omega_s,\omega_i\}$ space.  (b) Singles spectrum for a pump frequency which
satisfies $\omega_p>\omega_{zd}$. (c) Similar to (a), in the
non-degenerate pumps regime. (d) Singles spectrum for non degenerate
pumps such that $(\omega_1+\omega_2)>2\omega_{zd}$.}
\label{CONTspot}
\end{figure*}

Here we concentrate on the second, non-trivial solution. To this
end, it is instructive to consider a representation of the joint
spectrum (see Eq.(\ref{JSA1})) in $\{\delta_s,\delta_i\}$ (or
$\{\omega_s,\omega_i\}$) space. In Fig.~\ref{CONTspot}, we present
such a plot for a particular fiber (see figure caption), where we
have not, in contrast with the preceding analysis, resorted to a
truncated Taylor series description.  Here, the energy conservation
delta function is represented by a straight line with negative unit
slope, the trivial branch corresponds to a line with unit slope and
the non-trivial branch is represented by a curve which near
$\delta_s=\delta_i=0$ is elliptical. The spectral width of each of
the two phase matching branches is inversely proportional to the
fiber length (for graphical clarity in our plots we have assumed a
short fiber). We have also overlapped a plot of the contour defined
by  $\Delta k=0$, shown by the thin blue line, which corresponds to frequency
pairs which yield perfect phase matching. Photon pairs are generated
in areas of $\{\delta_s,\delta_i\}$ space where any of the two phase
matching branches meet the energy conservation locus.

Fig.~\ref{CONTspot}(a) shows an example of the joint spectral
intensity $|F\left(\omega_s,\omega_i \right)|^2$ for degenerate
pumps, obeying $\omega_p>\omega_{zd}$ (for which the energy conservation delta function is presented by the thick black, solid line), leading to a singles spectrum
$|f(\omega)|^2$ given by Eq.~\ref{espindiv} with three peaks as
shown in Fig.~\ref{CONTspot}(b); while the outer peaks correspond to
the nontrivial branch, the central peak corresponds to the trivial
branch which is centered at the pump frequency.  Note that if the
degenerate pump frequency is made equal to $\omega_{zd}$, the two
non-trivial peaks merge into a single peak, which then overlaps
with the trivial peak. The long-segment dashed line with negative
unit slope, represents the resulting energy conservation locus,
tangent to the phasematching curve at
$\omega_s=\omega_i=\omega_{zd}$. If the degenerate pump frequency is
tuned down further to values such that $\omega_p<\omega_{zd}$
(short-segment dashed line) then the non-trivial branch becomes
inaccessible, leaving only the trivial contribution. Note that if
the concavity of the phasematching ellipse is reversed (see below),
then spectrally-distinct trivial and non-trivial contributions are
observed for $\omega_p<\omega_{zd}$ instead.

An analysis of Eq.(\ref{dkdeg}) reveals that the non-trivial phase
matching branch, to fourth order in the phasemismatch, has an
elliptical locus in $\{\omega_s,\omega_i\}$ space.  Note that our
Taylor series description is valid only in the vicinity of the zero
group velocity dispersion frequency; for large enough detunings from
this frequency, the phase matching locus will deviate from the
elliptical shape. Note also that for the degenerate pump case, the
non-trivial phase matching characteristics are independent of the
pump field.  In this case, the phase matching curve is tangent at
the point given by $\omega_s=\omega_{zd}$ and $\omega_i=\omega_{zd}$
(\emph{i.e.} $\delta_s=\delta_i=0$) to a line with negative unit
slope. The curvature at this point is proportional to the ratio
$k^{(4)}/k^{(3)}$; thus, if this ratio is positive the concavity is
oriented towards negative values of $\delta_{+}$, while if the ratio
is negative, the concavity is oriented in the opposite direction.

\subsection{Phase matching properties for non-degenerate pumps}

Let us now restrict our attention to a SFWM process in the
non-degenerate pumps regime, i.e. $\omega_{1} \neq \omega_{2}$. For
negligible self/cross phase modulation, an analysis of
Eq.(\ref{deltakcw}) shows the existence of three distinct
phasematching branches, two of them trivial and one non-trivial. It
is straightforward to verify by direct evaluation of Eq.~(\ref{Dk})
in the low pump power limit, that frequency pairs on the straight lines
$\delta_i=\delta_s\pm \delta_{p-}$ (where
$\delta_{p-}=\delta_1-\delta_2$) fulfil perfect phasematching. These
are trivial phasematching branches; indeed, together with energy
conservation, they can be shown to be centered at:
i)$\omega_s=\omega_{1}$, $\omega_i=\omega_{2}$ and
ii)$\omega_s=\omega_{2}$, $\omega_i=\omega_{1}$.  In order to
describe the non-trivial branch, we initially restrict our attention
to a small spectral vicinity around $\omega_{zd}$ for which second
and higher-order terms in Eq.~\ref{deltakcw} can be neglected.  In
this case, from Eq.~\ref{deltakcw} it may be shown that frequency
pairs on the line $\delta_i=-\delta_s+2(2 \gamma P- \delta
k_0)/\delta k_1$ fulfil perfect phasematching.  The fourth order
term in Eq.~\ref{deltakcw} (as well as higher order terms not
included in our analysis) has the effect of adding curvature to this
non-trivial phasematching branch.

In order to illustrate this discussion, Fig.~\ref{CONTspot}(c) shows
a representation of the joint spectral intensity, for the same fiber
as in Fig.~\ref{CONTspot}(a), but now in the non-degenerate pumps regime (with $\lambda_{1}=2\pi c/\omega_{1}=0.5400\mu$m and
$\lambda_{2}=2\pi c/\omega_{2}=0.7000\mu$m). Note that here we have
not resorted to a truncated Taylor series description. The energy
conservation delta function is represented by a thick black, straight line with
negative unit slope, while the trivial branches are represented by
two straight lines with unit slope.  The non-trivial branch is the
curved line. As in the degenerate pump case, the spectral width of
each of the three phase matching branches is inversely proportional
to the fiber length. We have also overlapped a plot of the contour
defined by $\Delta k=0$, shown by the thin blue line. Photon pairs are generated
in areas of $\{\delta_s,\delta_i\}$ space where any of the three
phase matching branches meet the energy conservation locus.

The joint spectral intensity, represented by Fig.~\ref{CONTspot}(c),
leads to a singles spectrum with four peaks as shown in
Fig.~\ref{CONTspot}(d). While the outer peaks correspond to the
nontrivial branch, the two inner peaks correspond to the trivial
branches. In the case of non-degenerate pumps, depending on the pump
configuration, it is possible to observe: i)two non-trivial peaks
spectrally distinct from two trivial peaks (as in
Fig.~\ref{CONTspot}(d)), ii)a single non-trivial peak spectrally
distinct from two trivial peaks (if the energy conservation locus
meets the non-trivial branch tangentially), and iii)two trivial
peaks only (if the energy conservation locus does not meet the
non-trivial branch).

\subsection{Effect of self/cross phase modulation}

\begin{figure}[t]
\begin{center}
\centering\includegraphics[width=7.5cm]{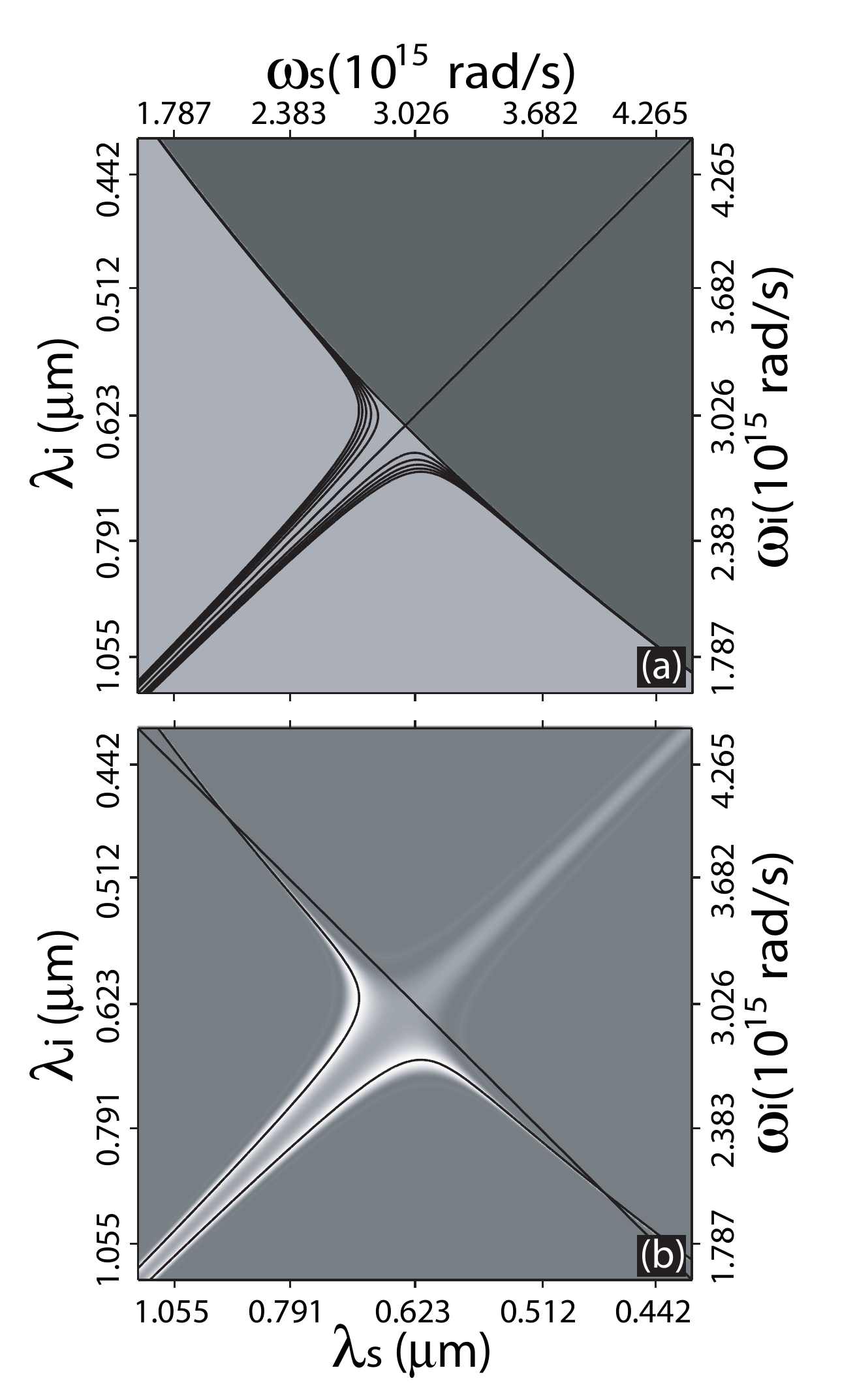}
\end{center}
\par
\caption{(a) Contours $\Delta k_{cw}(\omega_s,\omega_i)=0$ for the
following peak powers: $P=200$W, $P=400$W, $P=600$W, $P=800$W and
$P=1000$W (we have used large peak powers for graphical clarity).
The background shading represents the sign of the phasemismatch in
the low pump power limit $(\Delta k_{P=0})$; the dark areas
correspond to $\Delta k_{P=0}<0$. (b) Representation of the joint
spectral intensity for degenerate pumps with a fiber length of $L=3$cm
and a peak pump power of $P=1000W$; we have assumed values for $r$ and $f$
as specified in the caption of Fig.~\ref{CONTspot}.}
\label{Fig:CONTpot}
\end{figure}

Let us now consider the effect of the self/cross-phase modulation
term in the degenerate pumps regime (a similar behavior to that to be
described here would be observed for the non-degenerate pumps regime), which becomes important for a sufficiently high pump peak
power. For a given pump peak power, this term is an additive
constant to the phasemismatch at zero power $\Delta k_{P=0}$. Thus,
a contour diagram in $\{\delta_s,\delta_i\}$ space formed by the
contour $\Delta k=0$ for different power levels, represents a
contour map of the function $\Delta k_{P=0}$. Upon increasing the
pump peak power, the phase matching contour shifts towards areas of
$\{\delta_s,\delta_i\}$ space characterized by higher values of
$\Delta k_{P=0}$.  As an illustration, Fig.~\ref{Fig:CONTpot}(a)
shows, for a given fiber (see figure caption), the effect on the
phasematching contour of increasing the pump power.  In the
background, dark areas correspond to  $\Delta k_{P=0} \le 0$ and
light areas correspond to $\Delta k_{P=0}\ge 0$. It is clear that as
the pump power level is increased, the phase matching contour shifts
from the dark-light interface into the light areas. How much it
shifts at a particular point on the curve depends on how slowly
$\Delta k_{P=0}$ changes with frequency at that point. In general
terms, for frequencies generated far from the pump frequency, e.g.
on the non-trivial branch, this shift is much less pronounced than
for frequencies generated near the pump frequency.

For the trivial phasematching branch and degenerate pumps, the
self/cross phase modulation term leads to the suppression of perfect
phasematching in areas of $\{\omega_s,\omega_i\}$ space where
$\Delta k_{P=0}\leq0$ (indicated with dark shading in
Fig.~\ref{Fig:CONTpot}(a)). In contrast, in areas characterized by
$\Delta k_{P=0}\geq0$, the self/cross-phase modulation term splits
the trivial branches into two parallel branches. In
Fig.~\ref{Fig:CONTpot}(b) we show a representation of the
corresponding joint spectral intensity.  The plot in shades of gray
represents the phasematching function $|\mbox{sinc}[L \Delta
k_{cw}(\omega_s,\omega_i)/2]|^2$, which is maximized, with $\Delta
k_{cw}=0$, for frequency pairs on the curved black lines; the
straight black line represents the energy conservation locus.  Note
that while the dark-shaded area in Fig.~\ref{Fig:CONTpot}(a) can be
accessed with a pump in the normal dispersion ($k^{(2)}>0$) regime,
the light-shaded area can be accessed with a pump in the anomalous
dispersion ($k^{(2)}<0$) regime. The splitting of the trivial branch
in the anomalous-dispersion regime is a manifestation of  modulation
instability.  This phenomenon, which is equivalent to
four-wave mixing phasematched by the self/cross-phase modulation,
involves the appearance of two spectral sidebands, placed
symmetrically around the pump\cite{agrawal2007}.  In the time
domain, and in the stimulated regime, this leads to an ultrafast
modulation with period $4 \pi/\delta \Omega$ (where $\delta \Omega$
is the spectral separation between the two sidebands).

\section{Conditions for the generation of ultra-broadband photon pairs}\label{condiciones}

In this section, we derive conditions which guarantee the emission
of highly broadband photon pairs by spontaneous four wave mixing
utilizing narrowband pumps, in both the degenerate and
non-degenerate pumps regimes.  We will carry out the derivation of
these conditions in the low pump power limit; in practice, these
conditions are applicable for peak pump powers up to a few watts.
From Eq.(\ref{JSA1}) it is clear that SFWM photon pair emission
occurs at signal and idler frequency pairs
$\{\omega_{s},\omega_{i}\}$ which satisfy energy conservation, i.e.
$\omega_{s}+\omega_{i}=\omega_{1}+\omega_{2}$, and which in addition
satisfy phasematching, i.e. $\Delta
k_{cw}(\omega_{s},\omega_{i})=0$.  In order to obtain  a large
generation bandwidth, we must therefore engineer the phasematched
region in the joint frequency space $\{\omega_s,\omega_i\}$  to
maximize its overlap with the energy conservation locus, which is
itself a straight line parallel to the $\delta_{-}$ axis.   This
translates into a condition on the orientation, curvature and
location, of the phasematching contour $\Delta
k_{cw}(\omega_{s},\omega_{i})=0$.

Concretely, we expect broadband photon pair generation centered at
signal and idler frequencies $\{\omega_{so},\omega_{io}\}$ if four
conditions are satisfied by these frequencies: i) they satisfy
perfect phasematching, \emph{i.e.} $\Delta
k_{cw}(\omega_{so},\omega_{io})=0$,  ii) they obey energy
conservation, \emph{i.e.} the energy conservation locus contains
this frequency pair,  iii) the phasematching contour, which contains
this frequency pair [as guaranteed by i)], is oriented parallel to
the $\delta_-$ axis, and iv) the curvature of the phasematching
contour vanishes at this frequency pair.  Let us note that all
degenerate frequency pairs $\omega_{so}=\omega_{io}$ (i.e. which are
contained by the line $\delta_{-}=0$) fulfil the orientation
requirement iii); indeed, from Eq.~\ref{deltakcw} it may be shown
that the phasematching contour fulfils ($d \delta_{+}/d
\delta_{-})|_{\delta_{-}=0}=0$.  We will therefore restrict
attention to SFWM geometries centered at a degenerate frequency
pair.  Constrained by $\delta_{-}=0$, the attainment of
phasematching [condition i) above] is guaranteed if, in addition,
$\delta_{+}=0$ as can be verified from Eq.~\ref{deltakcw} (note that
$\delta k_0$ vanishes for $\delta_{+}=\delta_{-}=0$). Energy
conservation [condition ii) above] can then be satisfied for
$\delta_{+}=\delta_{-}=0$ if $\delta_{p+}=0$ (see
Eq.{\ref{jsadpmas}), which in turn leads to two possible scenarios
involving perfect phasematching at $\omega_s=\omega_i=\omega_{zd}$:
i) degenerate pumps (DP) with $\omega_1=\omega_2=\omega_{zd}$ (or
$\delta_1=\delta_2=0$), and ii) non-degenerate pumps (NDP) with
$\omega_2=2 \omega_{zd}-\omega_1$, where $\omega_1,\omega_2 \neq
\omega_{zd}$ (or $\delta_1=-\delta_2$ with $\delta_1,\delta_2 \neq
0$).

Conditions i) through iii) above can be fulfilled for any given
fiber; they determine the pump and central generation frequencies
which lead to the optimum SFWM bandwidth, for a specific fiber.  The
condition on the curvature [iv) above], on the other hand, helps us
to select the fiber core radius, for a given air-filling fraction (or if our analysis were applied to standard step-index fiber, it helps to select the core radius for a given index contrast).
In order to analyze the fulfilment of this condition, it is helpful
to define $C=|\delta_{+}''(1+\delta_{+}'^2)^{-3/2}|$,  where $'$
denotes a derivative with respect to $\delta_{-}$ evaluated at
$\delta_{-}=0$. This quantity represents the curvature of the
phasematching contour at the origin of the generated frequencies
$\{\delta_{+},\delta_{-}\}$ space.  It is possible to verify that
the curvature associated with the non-trivial branch, in the
low-power limit, is given by $C=|k^{(4)}/(12 k^{(3)})|$.  Thus, the
curvature is eliminated at radii for which $k^{(4)}\ll k^{(3)}$; in
general, as will be discussed below, for each $f$, one value of $r$
exists such that $k^{(4)}=0$.  The role of $k^{(4)}$ is illustrated
in Fig.~\ref{fig: contbroadbanddeg} for the degenerate pumps regime,
where we present phasematching contours, represented by solid lines,
in a generated frequencies (vertical axis) vs pump frequency
(horizontal axis) diagram.  Here, we express the generated
frequencies as detunings from the pump frequency
$\Delta=\omega-\omega_p$; the top half of the diagram corresponds to
the signal photon, while the bottom half corresponds to the idler
photon.   The three panels show the phasematching contour for a PCF
with $f=0.5$, for three different values of the core radius, for
which: i)$k^{(4)}<0$ (with $r=0.9023\mu$m) in panel (a),
ii)$k^{(4)}=0$ (with $r=0.7059\mu$m) in panel (b) and iii)
$k^{(4)}>0$ (with $r=0.5953 \mu$m) in panel (c). Note that in all
three cases, conditions i) through iii) from Sec.~\ref{condiciones}
are fulfilled.  It is clear that in panel (b), the phasematching
contour has an essentially vertical character [due to the fulfilment
of condition iv)], indicating that even a narrow pump bandwidth is
capable of generating remarkably broadband photon pairs.

\begin{figure*}[ht]
\begin{center}
\centering\includegraphics[width=17cm]{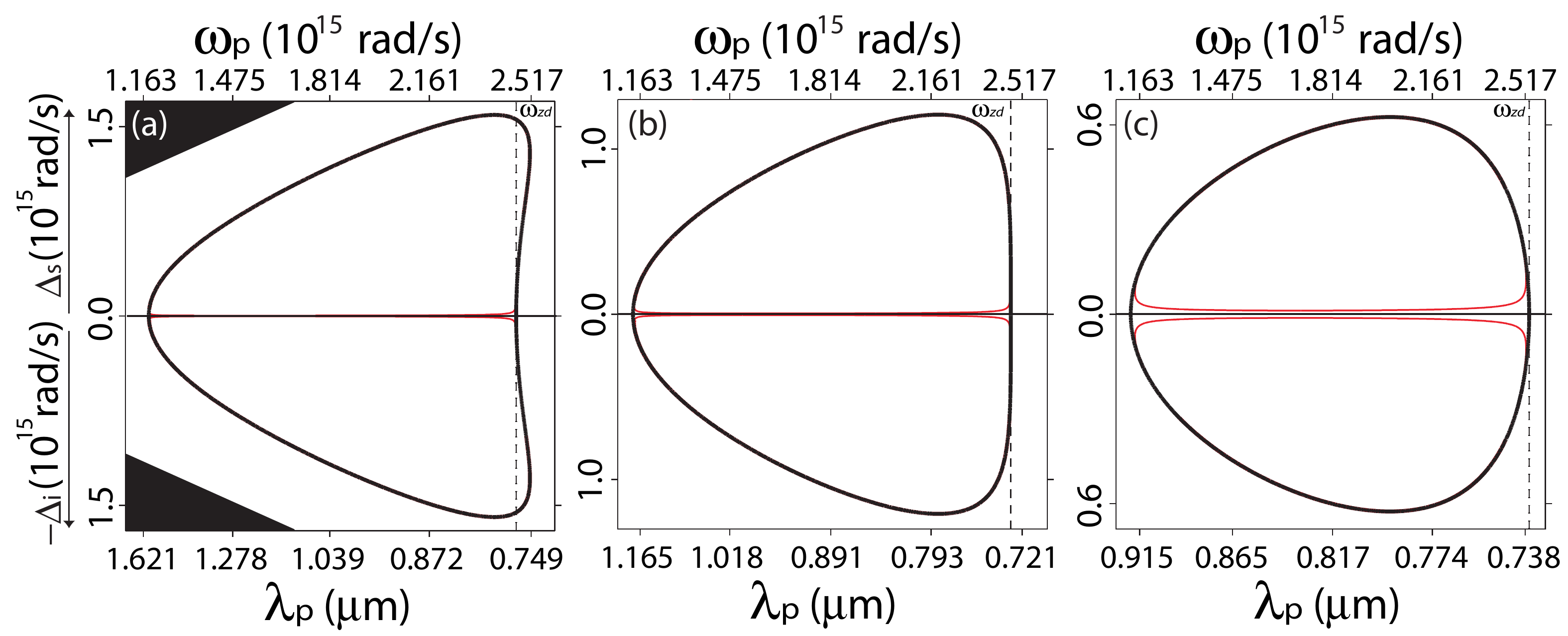}
\end{center}
\caption{(Color online) Phasematching [$\Delta k_{cw}(\omega_s,\omega_i)=0$]
contours for fibers with $f=0.5$ and: (a) $r=0.9023\mu$m, (b)
$r=0.7059\mu$m and (c) $r=0.5953 \mu$m. While the thin red curves
represent the phasematching contours for a peak power of $P=5$W, the
overlayed thick black curves were computed in the low pump power limit.
Vertical dashed lines mark the zero dispersion frequency on the pump
frequency axes. Note that for $r=0.7059\mu$m, the contour is
essentially vertical at $\lambda_p=\lambda_{zd}=0.7292\mu$m,
yielding a wide generation bandwidth of signal and idler photon
pairs.  The areas shaded in black represent non-physical regions
(where $\omega_s$ or $\omega_i$ would have to be negative to satisfy
energy conservation).}
\label{fig: contbroadbanddeg}       
\end{figure*}

Thus, for a source geometry which satisfies conditions i) through
iv) above, the phasematching contour is a straight line around the
zero dispersion point $\delta_s=\delta_i=0$, which in addition
coincides with the energy conservation locus.  For frequency pairs
sufficiently removed from $\delta_s=\delta_i=0$, the phasematching
contour will depart from a straight line due to terms $k^{(n)}$ with
$n \ge 5$.  Therefore, the attainable phasematching bandwidth is
limited by the magnitude of these dispersion coefficients.   Let us
note that a fiber characterized by a lower air filling fraction $f$
involves a lower core-cladding dielectric contrast, and consequently
the dispersive effects due to the waveguide contribution tend to be
weakened. Such a weaker dispersion leads to reduced dispersion
coefficients $k^{(n)}$, which in turn limits the departure from a
straight line of the phasematching contour.   Therefore, lower values
of $f$  permit greater SFWM bandwidths.

The process of SFWM leads to a remarkable symmetry between the two
pumps configurations (DP and NDP) discussed above.   Note that the
two-photon state is determined by the phasemismatch function $\Delta
k_{cw,sing}(\omega)$ (see Eq.~\ref{delksing}); thus, if we can show
that the DP and NDP scenarios lead to identical phasemismatch
functions, this would imply that the two-photon states in these two
scenarios are themselves identical. To this end, it is instructive
to calculate the phasemismatch function difference between these two
scenarios

\begin{eqnarray}
\label{dif_dk} &\Delta k_{cw,sing}^{NDP}(\omega)-\Delta
k_{cw,sing}^{DP}(\omega)= k(\omega_{1})+k(2
\omega_{zd}-\omega_{1})\nonumber\\&-2 k(\omega_{zd})=\Delta
k^{NDP}_{cw,sing}(\omega_{zd};P=0)=\delta k_0.
\end{eqnarray}

As indicated in Eq.~\ref{dif_dk}, the phasemismatch function
difference between the DP and NDP cases is equal to the NDP
phasemismatch, for vanishing pump power and evaluated at
$\omega=\omega_{zd}$, which is in turn equal to $\delta k_{0}$ (see
Eq.~\ref{deltakcw}).  Thus, provided that perfect phasematching is
achieved at degenerate signal and idler frequencies coinciding with
$\omega_{zd}$, $\Delta k_{cw,sing}^{NDP}(\omega)=\Delta
k_{cw,sing}^{DP}(\omega)$, and therefore the JSA functions, as given
by Eq.~\ref{JSA1}, for these two scenarios are in fact identical.
Note that this symmetry is lost for sufficiently large pump
bandwidths. Note, also, that this symmetry applies to the JSA, i.e.
to the two-photon state, but not to the underlying phasematching
function. In section~\ref{interferencia} we discuss an interference
effect, which arises due to identical outcomes obtained by two indistinguishable pathways, when pumping with DP and NDP pumps
simultaneously.

For a given fiber (characterized by the values for $r$ and $f$), the
singles spectrum $|f(\omega)|^2$ can be computed using
Eq.~\ref{espindiv}, for a pump frequency such that conditions i)
through iii) above are satisfied, in either the DP or NDP regimes.
Given the symmetry discussed above, these two regimes give identical
SFWM joint, as well as singles, spectra.   Solid curves in
Fig.~\ref{ancholamzd}(a) show the resulting optimum  full width at
half maximum singles bandwidth as a function of the core radius,
where each curve is computed for a fixed value of the air-filling
fraction.  Here we have assumed a fiber length of $L=25$cm, strictly
monochromatic pumps and a vanishingly low pump peak power (we
consider the effect of realistic pump power levels in the next
section). Fig.~\ref{ancholamzd}(b) shows the zero dispersion
frequency, which if conditions i) through iii) are satisfied
corresponds to the central SFWM frequency, plotted vs the core
radius for different values of the air filling fraction.  These
plots show that while spectra centered in the infrared tend to be
broader than those centered in the visible, there is a considerable
flexibility for choosing the central emission frequency while
maintaining a large emission bandwidth (as compared to an approach
based on PDC with $\chi^{(2)}$ crystals). The spectra from which the
curves in Fig.~\ref{ancholamzd}(a) were calculated typically
include a main emission mode involving an essentially flat spectrum
and in some cases pairs of peaks around the main emission mode. For
solid-line curves in Fig.~\ref{ancholamzd}(a), the valleys between satellite peaks and the main
emission mode do not drop below $0.5$ of the maximum value. For
certain radii, however, the spectrum exhibits pairs of well-defined
satellite peaks around the main emission mode, resulting in voids in
the spectrum.  The bandwidth indicated in Fig.~\ref{ancholamzd}(a)
by dashed-line curves is calculated from the outer-most slopes
associated with these satellite peaks.   It is important to point out that in
the case of the NDP regime, these satellite peaks are in fact
centered at the pump frequencies, and are due to trivial
phasematching.  Fig.~\ref{fig: anchoairfillfra} shows a curve
similar to that in Fig.~\ref{ancholamzd}(a) for $f=0.1$ including
the complete branch associated with satellite peaks.  As can be
seen, there are fiber core radii for which there are in fact more than one pair of satellite peaks.  This is clear, for example, for the core radius labelled as C, for which there are two pairs of satellite peaks.

In Fig.~\ref{ancholamzd}(a), each solid-line curve is interrupted
towards smaller radii, at the radius for which the zero dispersion
frequency no longer exists [conditions i) through iii) assume the
existence of a zero dispersion frequency].  It is apparent from the
figure that the generation bandwidth tends to be larger for smaller
air filling fractions.  It is also apparent that for a given
air-filling fraction, the attainable bandwidth exhibits a peak near
the core radius $r_{k^{(4)}=0}$ for which the condition
$k^{(4)}=0$ is fulfilled (shown as red dots).  Note that the exact radius
yielding the maximum bandwidth is slightly larger than
$r_{k^{(4)}=0}$.  This is due to the influence of higher-order
dispersive terms.  In particular, it turns out that a small degree
of curvature, with $k^{(4)} \neq 0$, can enhance the attainable
bandwidth, if $k^{(4)}$ and $k^{(6)}$ have opposite signs.  These
parameters do in fact have opposite signs over a small range of
radii larger than $r_{k^{(4)}=0}$, explaining the shift of the peak
from $r=r_{k^{(4)}=0}$.

\begin{figure}[h]
\begin{center}
\centering\includegraphics[width=7.5cm]{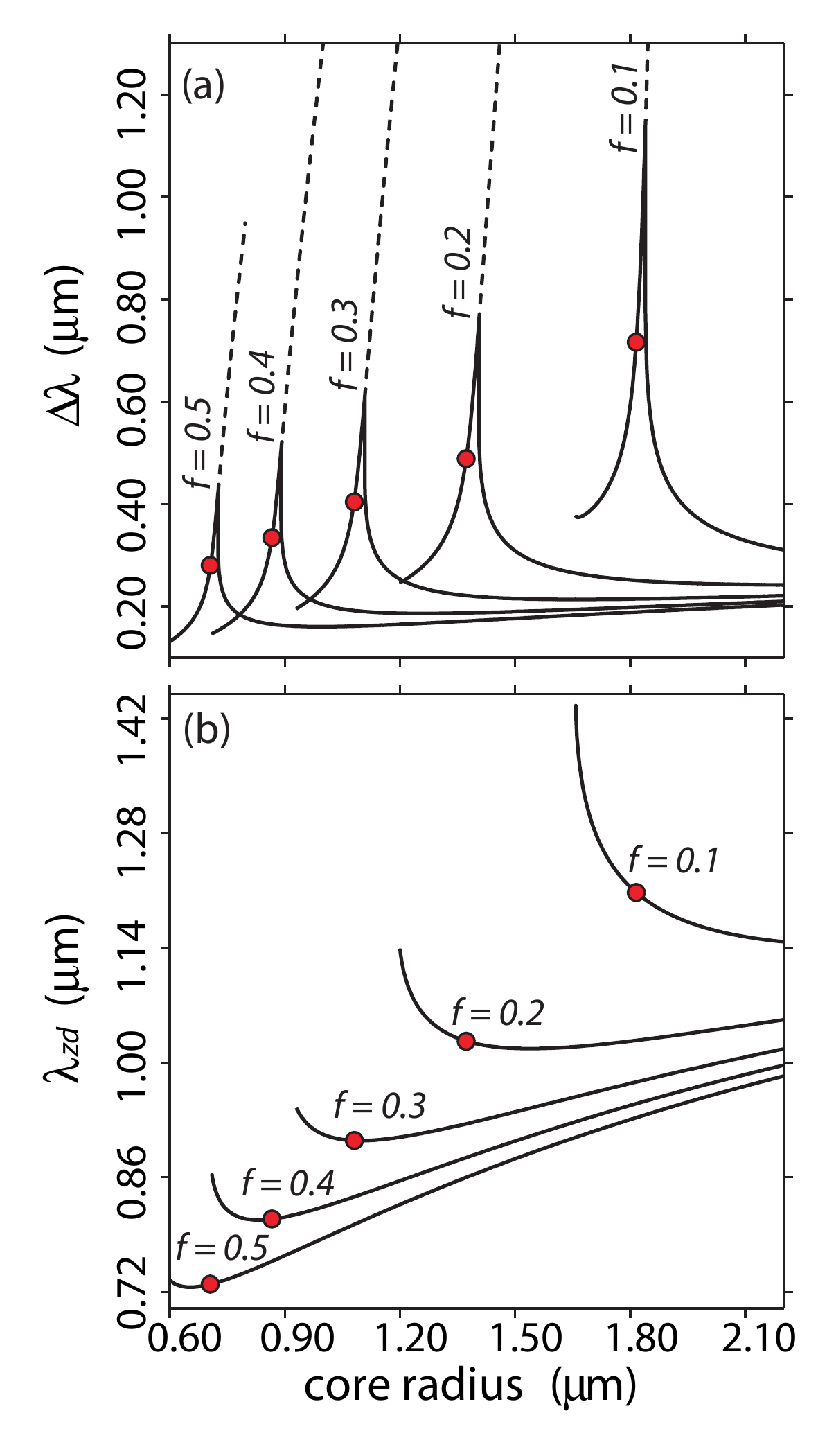}
\end{center}
\caption{(Color online) (a) Solid lines: FWHM bandwidth in the main emission mode
as a function of $r$ for different values of $f$. Dotted lines: FWHM
bandwidth calculated from the outer slopes of satellite peaks; these lines are shown interrupted when the emitted spectrum reaches the edge of the
range of validity of the Sellmeier expansion for fused silica.  For
each $f$, the red dot corresponds to the radius for which
$k^{(4)}=0$ is fulfilled. (b) Zero dispersion frequency vs
core radius $r$, for different values of the air filling fraction
$f$.  Note that when conditions i) through iii) of
Sec.~\ref{condiciones} are satisfied, the zero dispersion frequency
coincides with the central SFWM emission frequency. }
\label{ancholamzd}       
\end{figure}

An interrupted dashed-line
curve indicates that the emitted spectrum reaches the edge of the
range of validity of the Sellmeier expansion used to compute
dispersion for fused silica.

It is possible to re-interpret the  phasematching contours in the
low pump peak power limit, of Fig.~\ref{fig: contbroadbanddeg}, as
non-degenerate pump frequencies in the vertical axis and degenerate
signal and idler central frequencies in the horizontal axis.  This
is inverted with respect to the original interpretation, i.e.
degenerate pumps in the horizontal axis and non-degenerate signal
and idler in the vertical axis.  The NDP pump frequencies which
satisfy  conditions i) through iii) above can now be determined by
setting the degenerate signal and idler frequency at the zero
dispersion frequency (denoted in the figure by a vertical dashed
line), and reading out, on the vertical axis, the required NDP pump
frequencies from the intersection of this vertical line with the
phasematching contour. Thus, the example in Fig.~\ref{fig:
contbroadbanddeg}(a) clearly shows the DP pump frequency coinciding
with $\omega_{zd}$ and also shows one set of NDP pump frequencies.
Fig.~\ref{fig: contbroadbanddeg}(b) shows that source designs for
which the phasematching contour curvature approaches zero ($k^{(4)}
\approx 0$), which yields a large generation bandwidth in the DP
regime, permit an infinite number of pump frequencies which can
function as NDP pump frequencies (throughout the portion of the
contour which is essentially vertical). This leads to the remarkable
conclusion that an identical two-photon state is expected for any
pair of pump frequencies amongst the possibly infinite set comprised
of: i)the DP pump frequency, ii) possibly one or more discrete NDP
frequency pairs, and iii) possibly a continuum of NDP pump frequency
pairs.

A key consideration in the design of fiber-based two-photon sources is possible contamination from spontaneous Raman scattering, which occurs over a bandwidth of $\sim 40$THz to the red from each pump spectral band.   Our treatment in this paper does not take into consideration Raman gain; within the Raman bandwidth, a full analysis must take into consideration the combined effects of SFWM and spontaneous Raman scattering and therefore our theory is not complete\cite{golovchenko90,lin07}.  However, since here we concentrate on broadband SFWM, where the generation bandwidth is much larger than the Raman bandwidth, our theory does adequately describe the overall two-photon state structure.    The need for Raman suppression, in order to ensure high-quality signal-idler correlations, suggests a useful application for the symmetry between the DP and NDP regimes.   For a configuration with an optimized SFWM bandwidth, we may pump at a pair of discrete NDP frequencies, which if sufficiently removed from the main emission mode, guarantees that this main emission mode is free from spontaneous Raman scattering.

Note that longitudinal fluctuations of the fiber properties may have an important effect on the properties of emitted light, particulary on the attainable bandwidth\cite{chen06}.   In practice, this will set a fabrication tolerance on the amplitude and characteristic period of the fluctuations.

\begin{figure}[h]
\begin{center}
\centering\includegraphics[width=7.5cm]{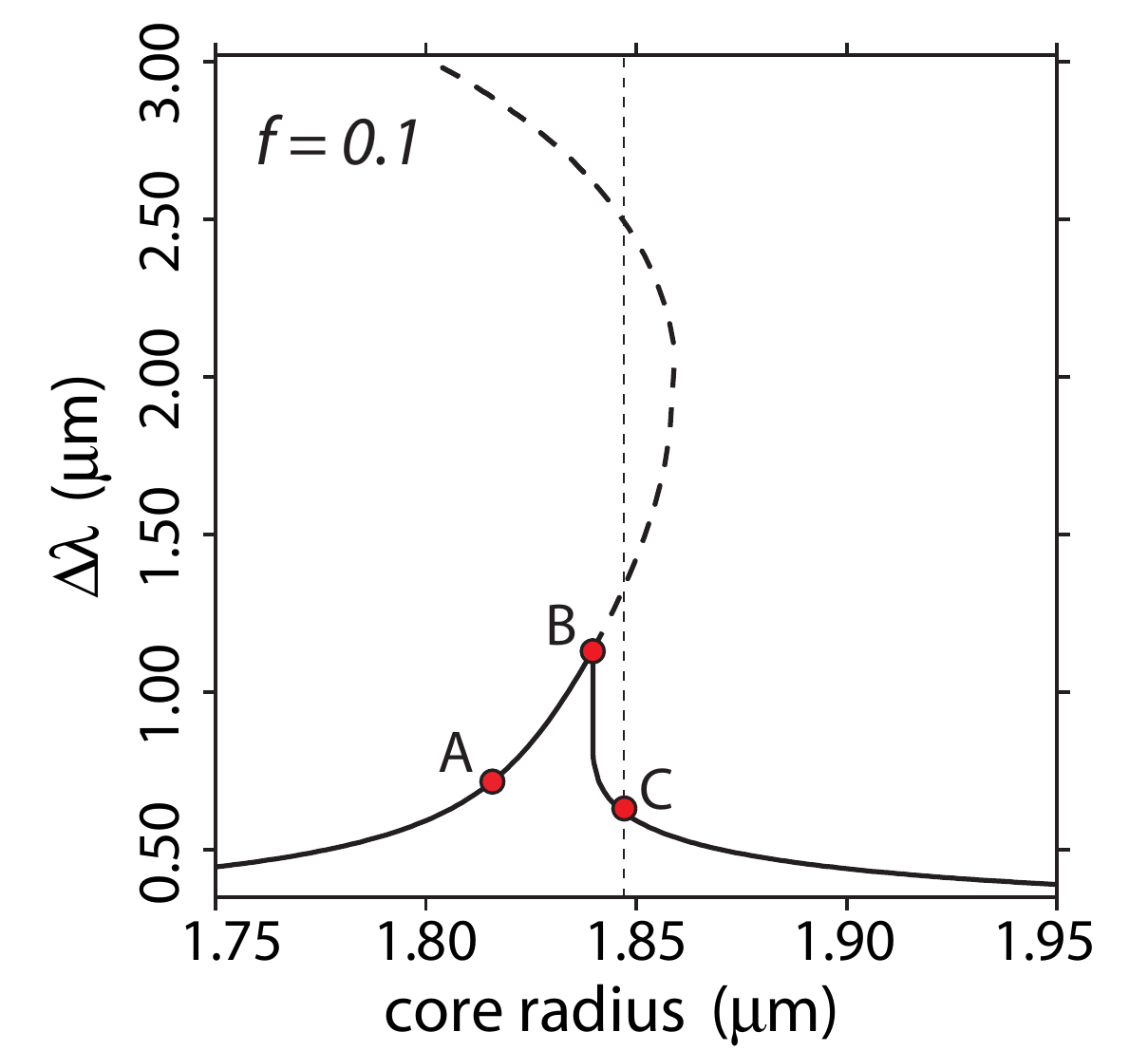}
\end{center}
\caption{(Color online) This figure is similar to Fig.~\ref{ancholamzd}(a), for the case $f=0.1$.   Here we have used a larger
plotting range to better represent the branch of the generated
bandwidth vs radius curve corresponding to the satellite peaks.  The
spectra associated with points A, B and C are discussed in
Sec.~\ref{casos}.}
\label{fig: anchoairfillfra}       
\end{figure}

\section{Ultra-broadband two-photon states: specific experimental designs}\label{casos}

In this section we present specific designs of fiber-based,
ultra-broadband photon-pair sources.  As we have noted in the
previous section, a small air-filling fraction tends to enhance the
resulting SFWM generation bandwidth.  Thus, for the specific
examples to be considered here, we assume fibers with $f=0.1$ (which
according to Ref.~\cite{Wong2005} represents the lower bound for the
range of validity of the step-index, effective-medium dispersion
model which we have employed). We have selected three different
source designs, involving radii labelled A, B and C in
Fig.~\ref{fig: anchoairfillfra}. For  source  A, the core radius is
chosen so that the condition $k^{(4)}=0$ is fulfilled (corresponding
to a vanishing phasematching contour curvature).  For source B, the
core radius is chosen so that we obtain the largest possible flat-spectrum
generation bandwidth possible for $f=0.1$ (as discussed above, this
maximum does not occur at the radius for which $k^{(4)}=0$ due to
the effect of higher-order terms).  For source C, we have selected
the smallest core radius for which the two satellite peaks become well
defined, i.e. such that the rate of emission reaches zero between
the main emission band and the satellite peaks. In what follows we
present emission spectra for these three cases. In all cases we will
assume a fiber length of $L=25$cm, a pump peak power of $P=5$W, a
nonlinear coefficient of $\gamma=70 \mbox{W}^{-1} \mbox{km}^{-1}$,
and a pump bandwidth of $\sigma=50$MHz, unless stated otherwise.

In Figs.~\ref{crotados} and \ref{casoA} we present the two-photon
state and singles spectrum obtained for an air-filling fraction
$f=0.1$ and for a core radius of $r=r_{k^{(4)}=0}=1.8162\mu$m (which
corresponds to point A in Fig.~\ref{fig: anchoairfillfra}). A number
of choices for the pump frequencies which satisfy conditions i)
through iii) above exist, and which therefore permit the largest
possible generation bandwidth for the fiber in question:
i)$\lambda_1=\lambda_2=\lambda_{zd}=2 \pi c/\omega_{zd}=1.2076\mu$m
in the DP regime, ii)$\lambda_{1}=0.7252\mu$m,
$\lambda_{2}=3.6070\mu$m (one individual set of NDP frequencies),
and iii) any pair of frequencies symmetrically displaced from
$\omega_{zd}$ within the range $1.1606\mu \mbox{m} \lesssim \lambda
\lesssim 1.2586\mu \mbox{m}$; these constitute a continuum of NDP
frequency pairs.  It should be stressed that given the symmetry
between the DP and NDP regimes, all of the above choices of pump
frequencies will result in a two-photon state with the same joint
spectrum $|F_{cw}(\omega_s,\omega_i)|^2$.

In Fig.~\ref{crotados}(a) we present a plot of the phasematching
function $|\mbox{sinc}(L \Delta k_{cw}/2)|^2$, in
$\{\delta_{+},\delta_{-}\}$ space, for the DP regime.  We have
overlapped a plot of the perfect phasematching contour $\Delta
k_{cw}(\delta_{+},\delta_{-})=0$ (for graphical clarity we have
interrupted this contour towards the narrow, outer branches). Note
that the trivial phasematching branch exhibits a power-induced
splitting into two parallel branches. Fig.~\ref{crotados}(b) shows
the corresponding plot of the phasematching function  in
$\{\delta_{+},\delta_{-}\}$ space, plotted for the NDP regime (where
we have chosen pump frequencies corresponding to the discrete NDP
frequency pair); specifically, we have chosen $\lambda_{1}=2 \pi
c/\omega_{1}=0.7252\mu$m and $\lambda_{2}=2 \pi
c/\omega_{2}=3.6070\mu$m. In both cases we have indicated with a
yellow vertical line the energy conservation locus; ideally, for a
large generation bandwidth the phasematching contour and the energy
conservation locus should coincide.   Remarkably, while the
phasematching functions in the DP and NDP regimes are considerably
different (as can be appreciated from the plots), the resulting two
photon states are identical as demanded by the symmetry derived in
the last section.  Indeed, the portion of the phasematching function
which coincides with the energy conservation locus may be seen to be
identical in these two cases. Fig.~\ref{casoA}(a) shows the joint
spectral intensity $|F_{cw}(\omega_s,\omega_i)|^2$, plotted as the
contour corresponding to $0.5$ of the maximum value (note that the
narrow diagonal width should be interpreted only schematically; the
lines used are much thicker than the actual spectral width). As
expected, from the symmetry discussed above, the joint spectral
intensity is identical for the DP and NDP regimes. Note that in
Fig.~\ref{casoA}(a) there is a main emission mode centered at the
zero dispersion frequency, and there are two satellite peaks, shown
circled, involving highly non-degenerate frequency pairs.

The large degree of spectral entanglement evident in
Fig.~\ref{casoA}(a), could in principle be computed through the
Schmidt number, $K$. However, because of the very large ratio of the large
to small diagonal widths, it is difficult for the sampling used to
suffice in giving a reliable numerical estimate of $K$.  Our
calculation gives us a lower bound $K>1.7\times10^3$ (where $K=1$
indicates a factorable, or un-entangled, state).

Fig.~\ref{casoA}(b)shows the singles spectrum, calculated from
Eq.~\ref{espindiv} (and therefore assuming ideal monochromatic
pumps), which once again is identical for the DP and NDP regimes.
This spectrum exhibits a main emission mode centered at
$\lambda_{zd}=2 \pi c/\omega_{zd}=1.2076\mu$m, with a remarkably
large FWHM bandwidth of $712.2$nm. The singles spectrum also
exhibits two satellite peaks centered at $\lambda_{1}=2 \pi
c/\omega_1=0.7252\mu$m and $\lambda_{2}=2 \pi
c/\omega_2=3.6070\mu$m, which in the NDP regime with two discrete
peaks [scenario ii) above] are centered at the pump wavelengths. The
emission bandwidth calculated from the outer slopes of these
satellite peaks is $2884.8$nm.  The fractional generation bandwidth
for the main emission mode $\Delta \omega/\omega_c$ where $\Delta
\omega$ is the FWHM bandwidth and $\omega_c$ is the central
frequency gives a value of $0.546$. This large generation bandwidth
leads to a small correlation time, of $\tau=3.4$fs, calculated as
the width of the signal/idler emission time difference distribution.

\begin{figure}[t]
\begin{center}
\centering\includegraphics[width=7cm]{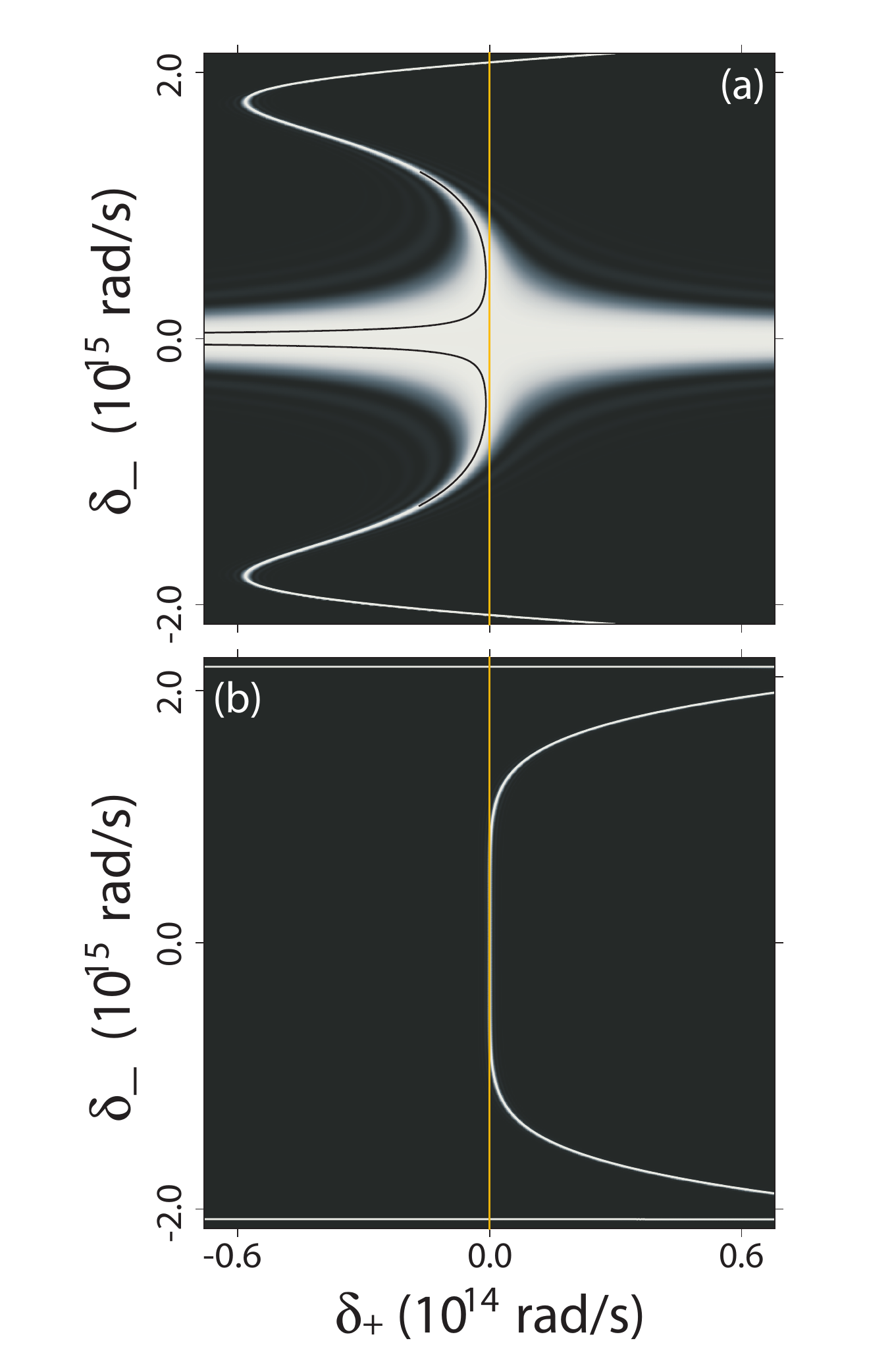}
\end{center}
\caption{(Color online) Representation of the joint spectral intensity in
$\{\delta_{+},\delta_{-}\}$ space obtained for a fiber corresponding
to point A in Fig.~\ref{fig: anchoairfillfra}.  Here we present a
plot of $|\mbox{sinc}[L \Delta k_{cw}(\omega_s,\omega_i)/2]|^2$ (in shades of gray), and
we also indicate the energy conservation locus as a vertical yellow line.
While (a) corresponds to the DP regime, (b) corresponds to the NDP
regime.}
\label{crotados}       
\end{figure}

\begin{figure}[t]
\begin{center}
\centering\includegraphics[width=7.38cm]{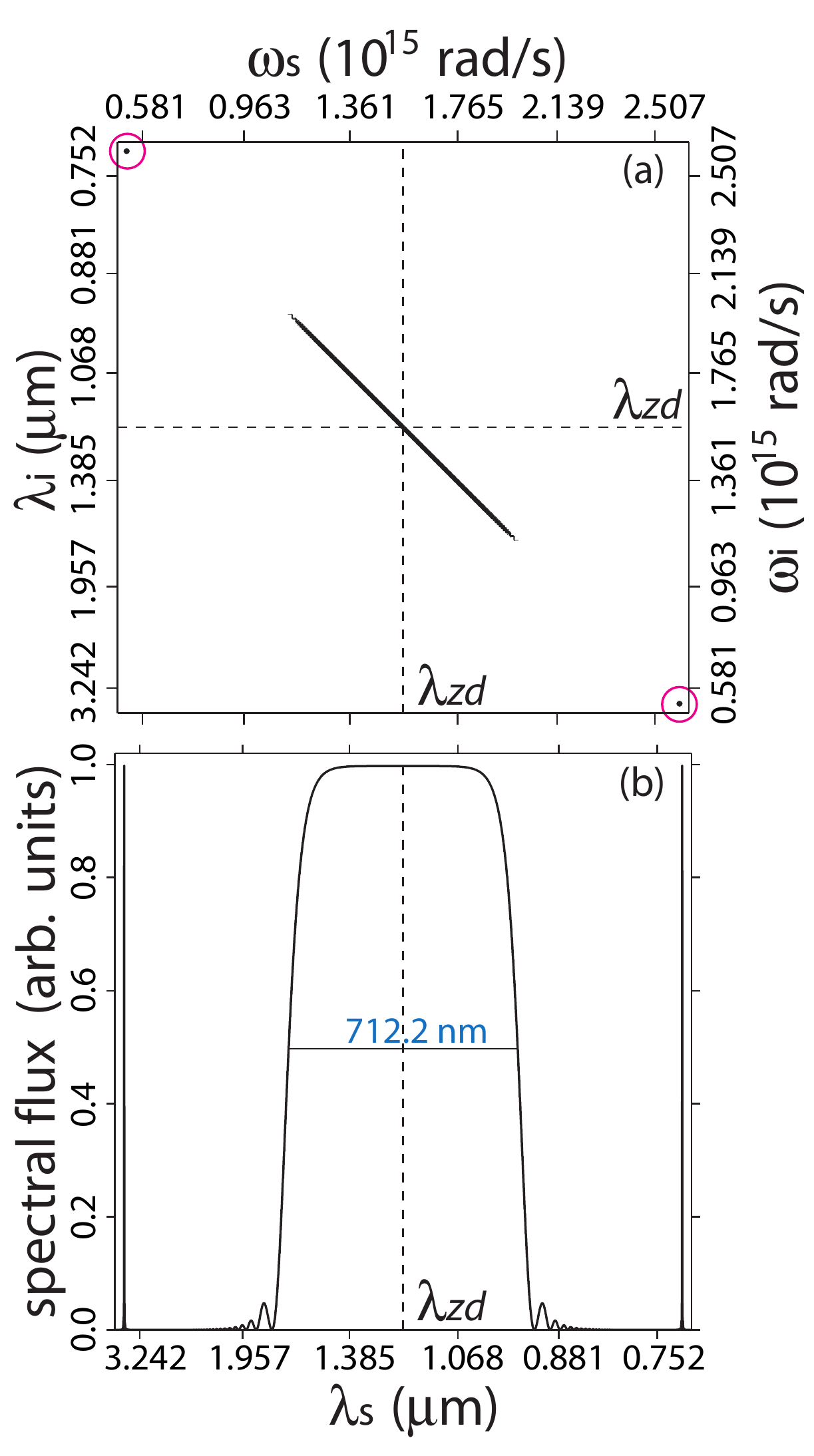}
\end{center}
\caption{(Color online) (a) Representation of the joint spectral intensity in $\{\omega_s,\omega_i\}$ space, for the
same parameters as in Fig.~\ref{crotados}.  Note that there are two
highly non-degenerate satellite peaks, shown circled. (b) Singles
spectrum corresponding to the joint spectral intensity in (a). }
\label{casoA}       
\end{figure}

In Fig.~\ref{casoB} we present the singles spectrum obtained for an
air filling fraction of $f=0.1$ and a core radius of $r=1.8402\mu$m
(corresponding to point B in Fig.~\ref{fig: anchoairfillfra}, which
yields the maximum possible flat-spectrum bandwidth for this air-filling fraction). As in the previous
example, a number of choices for pump frequencies which satisfy
conditions i) through iii) above exist.  These include i)
$\lambda_1=\lambda_2=\lambda_{zd}=2 \pi c/\omega_{zd}=1.1987\mu$m in
the DP regime, two discrete sets of NDP frequencies, at ii)
$\lambda_{1}=0.8683\mu$m, $\lambda_{2}=1.9349\mu$m and iii)
$\lambda_{1}=0.7306\mu$m, $\lambda_{2}=3.3367\mu$m, as well as iv)
any pair of frequencies symmetrically displaced from $\omega_{zd}$
within the range $1.1835\mu \mbox{m} \lesssim \lambda \lesssim
1.2145 \mu \mbox{m}$; these constitute a continuum of NDP frequency
pairs. Pump frequencies for scenarios i) and ii) are indicated in
the figure by vertical dashed lines. Note that the slight departure
from a flat central emission-mode is power-induced (an essentially
flat spectrum would be recovered for a lower pump peak power). This
source leads to a remarkably large generation bandwidth of $\Delta
\lambda=1142.3$nm, including a central mode and two adjacent
satellite peaks;  this corresponds to a fractional bandwidth of $\Delta \omega/\omega_c=0.794$.  Defining the bandwidth according to the outer
slopes of the outer-most satellite peaks, the generation bandwidth
becomes $2612.2$nm.   Note that if non-degenerate pump frequencies
are made to coincide with the first set of satellite peaks [scenario
ii) above], then the main emission mode remains free from
spontaneous Raman scattering; indeed the red rectangles in the
figure indicate the $\sim 40$THz spontaneous Raman scattering
bandwidth. The large bandwidth of the main emission mode leads to a
short correlation time of $\tau=2.5$fs.

\begin{figure}[t]
\begin{center}
\centering\includegraphics[width=6.8cm]{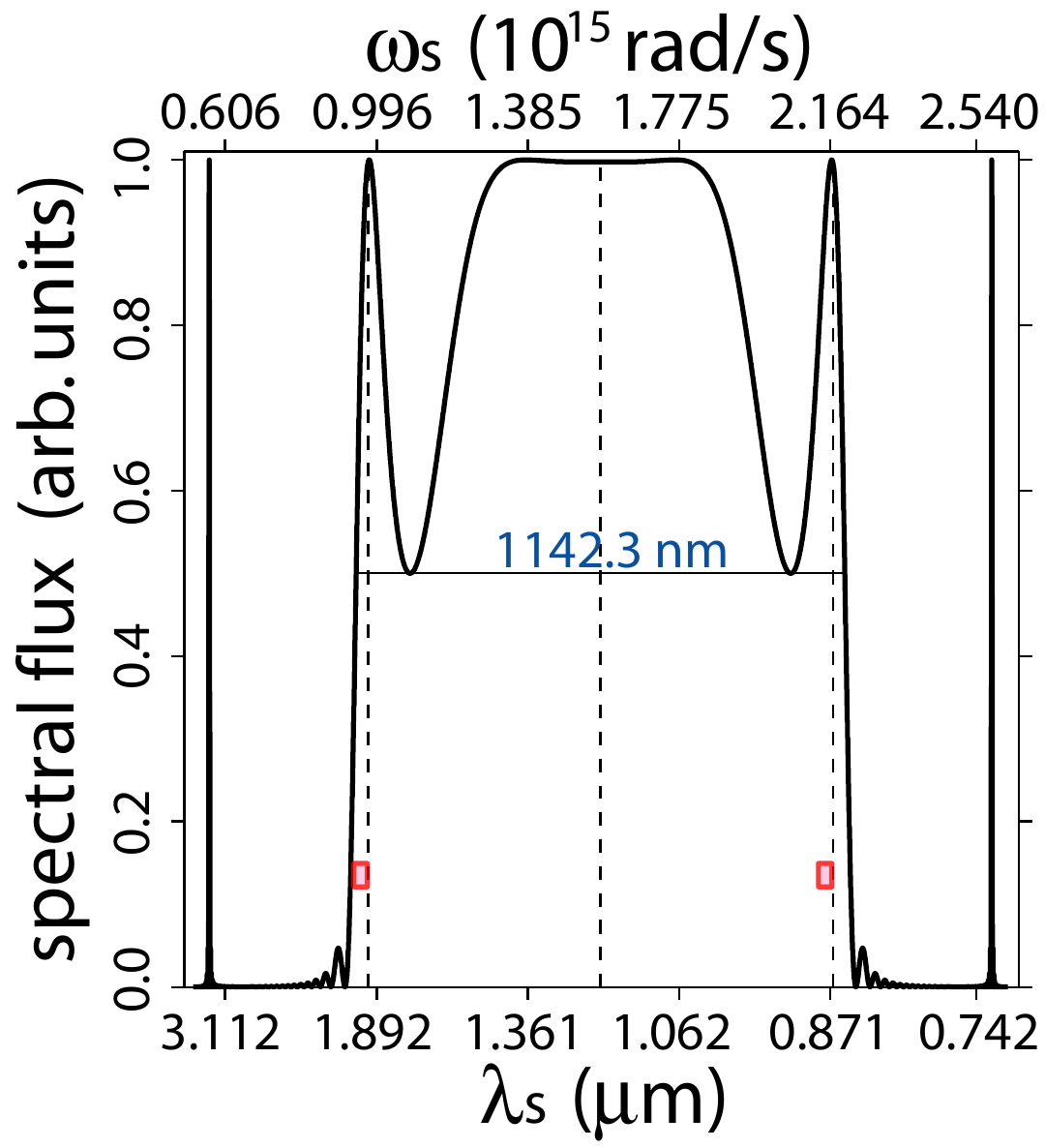}
\end{center}
\caption{(Color online) Singles spectrum corresponding to point B in Fig.~\ref{fig:
anchoairfillfra}.  The red rectangles represent the $\sim 40$THz
spontaneous Raman scattering bandwidth centered at the inner NDP
frequency pair. }
\label{casoB}       
\end{figure}

In Fig.~\ref{casoC} we present the singles spectrum obtained for an
air filling fraction of $f=0.1$ and a core radius of $r=1.8471\mu$m
(corresponding to point C in Fig.~\ref{fig: anchoairfillfra}), which
represents the smallest core radius for which the two satellite
peaks become well defined). As in the previous examples, a number of
choices for pump frequencies which satisfy conditions i) through
iii) above exist.  These include i)
$\lambda_1=\lambda_2=\lambda_{zd}=2 \pi c/\omega_{zd}=1.1964\mu$m in
the DP regime, two discrete sets of NDP frequencies, at ii)
$\lambda_{1}=0.8347\mu$m, $\lambda_{2}=2.1112\mu$m and iii)
$\lambda_{1}=0.7347\mu$m, $\lambda_{2}=3.2195\mu$m, as well as iv)
any pair of frequencies symmetrically displaced from $\omega_{zd}$
within the range $1.1821\mu \mbox{m} \lesssim \lambda \lesssim
1.2112\mu \mbox{m}$; these constitute a continuum of NDP frequency
pairs. Pump frequencies for scenarios i) and ii) are indicated in
the figure by vertical dashed lines.  As in case B, the slight
departure from a flat central emission-mode is power-induced, and
would be eliminated in the low pump power limit. This source leads
to a generation bandwidth of $\Delta \lambda=630.6$nm in the
central emission mode, corresponding to a fractional bandwidth of $\Delta \omega/\omega_c=0.495$ and to a correlation time of
$\tau=4.1$fs. Defining the bandwidth according to the outer slopes
of the inner-most satellite peaks, the generation bandwidth becomes
$1330.8$nm. Likewise, defining the bandwidth according to the outer
slopes of the outer-most satellite peaks, the generation bandwidth
becomes $2493.6$nm.

\begin{figure}[t]
\begin{center}
\centering\includegraphics[width=6.8cm]{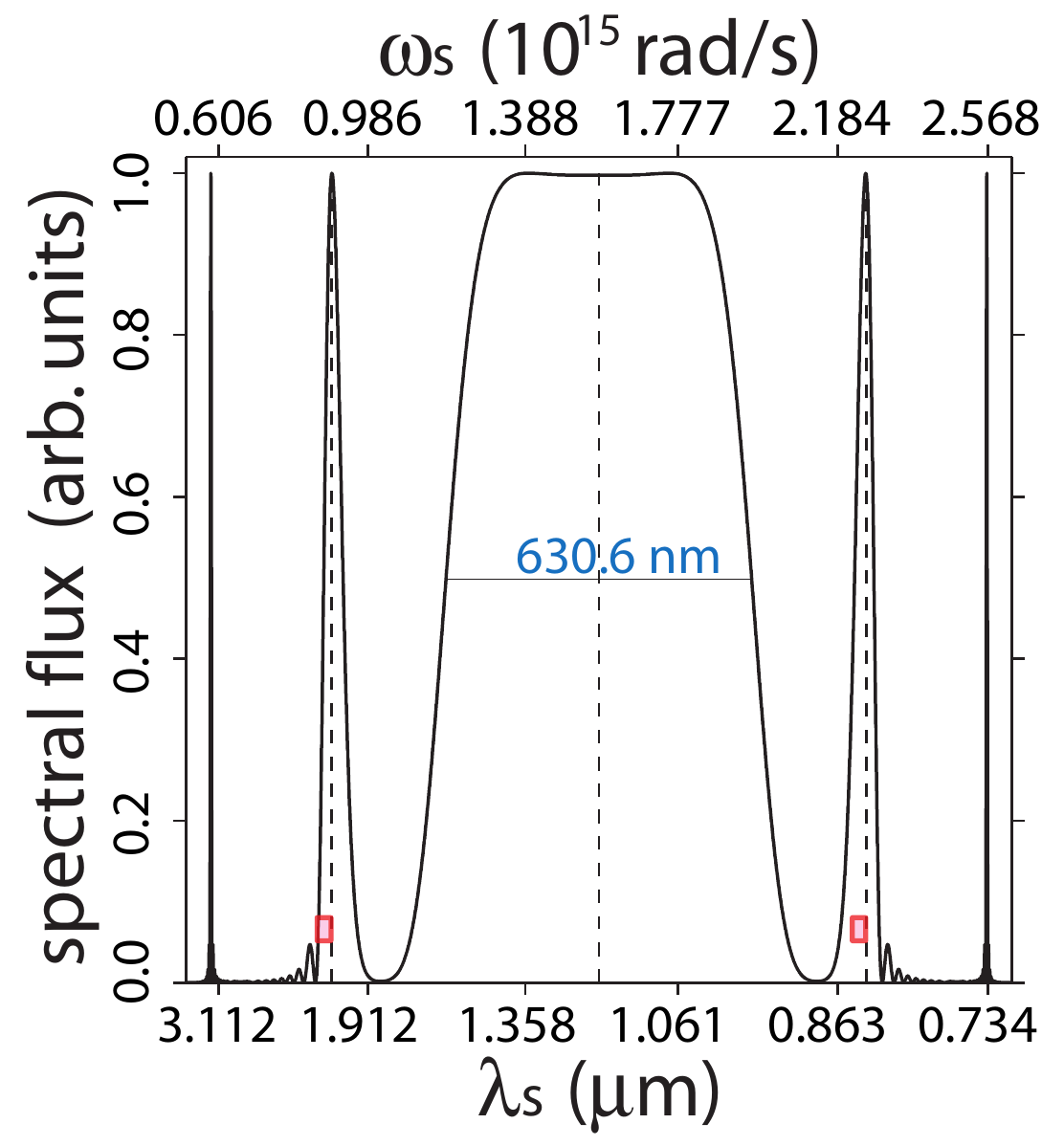}
\end{center}
\caption{(Color online) Singles spectrum corresponding to point C in Fig.~\ref{fig:
anchoairfillfra}.  The red rectangles represent the $\sim 40$THz
spontaneous Raman scattering bandwidth centered at the inner NDP
frequency pair. }
\label{casoC}       
\end{figure}

\section{Quantum interference in spontaneous four-wave
mixing}\label{interferencia}

In section~\ref{condiciones} we concluded that SFWM source designs
exist where the pump frequencies may be chosen from a certain set,
which may be finite or infinite, in such a way that any choice
within this set yields an identical two-photon state.  This set of
pump frequencies is composed of all those which lead to the
fulfilment of conditions i) through iii) of
section~\ref{condiciones}. It includes one DP frequency, coinciding
with the zero dispersion frequency, and can also include individual
pairs of NDP frequencies, as well as a continuum of NDP frequencies
centered at the zero dispersion frequency.  In this section we
explore the effect of pumping simultaneously at the frequencies
corresponding to two elements of this set, where coherence between
the various frequencies involved is assumed to exist.  In the
discussion which follows, we assume that these two elements
correspond to DP and NDP frequencies.  Because a specific photon
pair can then be created by any of two indistinguishable pathways,
we expect an interference effect to occur when we vary the relative
phase between these two sets of pump frequencies.  This interference
is such that the two individual pathways can interfere destructively
leading to the suppression of photon pair emission, or they can
interfere constructively. Variation of the relative phase between
the two sets of pump frequencies, then leads to a sinusoidal
modulation of the resulting rate of emission.

In order to analyze this interference effect, we assume a pump field
including three narrow spectral bands (e.g. with a rectangular
shape) centered at each of the DP and NDP frequencies.  It is
straightforward to verify that the total peak power available in
each of the DP and NDP regimes should be identical,  so that each
pump configuration leads not only to the same joint spectrum, but
also to the same rate of emission.  One way of attaining this
condition is if all pump spectral bands gave the same spectral
width, where in addition the DP amplitude is $\sqrt{2}$ higher than
those for the NDP pumps. Specifically, we assume that the pump
spectrum can be written as follows

\begin{align}
\label{pumprect}
\alpha(\omega)&=\alpha_0 [\mbox{rect}(\omega_{NDP,1}; \delta \omega)+\sqrt{2}e^{i \theta}\mbox{rect}(\omega_{DP}; \delta \omega) \nonumber \\
&+ \mbox{rect}(\omega_{NDP,2}; \delta \omega)],
\end{align}

\noindent where $\delta \omega$ is the bandwidth (assumed to be
small) for each of the pump spectral components, $\theta$ is the
phase introduced between the two pump configurations and
$\mbox{rect}(x;\delta_x)$ is unity for $-\delta_x/2 <x< \delta_x/2$
and zero otherwise.  In the monochromatic  limit where $\delta
\omega \rightarrow 0$ , each of the spectral bands may be
represented by a delta function and the joint spectral amplitude may
be computed analytically from Eq.~\ref{eq: JSA} yielding

\begin{equation}
\label{JSAinterf} F(\omega_s,\omega_i)=N' e^{i \theta} \cos \theta
F_{DP}(\omega_s,\omega_i),
\end{equation}

\noindent in terms of $F_{DP}(\omega_s,\omega_i)$ which represents
the joint spectral amplitude for either the DP or NDP regimes [which
are identical if conditions i) through iii) above are satisfied],
and where $N'$ represents a normalization constant. In
Eq.~\ref{JSAinterf} we have omitted a number of terms which are
negligible, within the spectral range of interest, for a sufficient
spectral separation between the DP and NDP spectral components. Note
that the non-linear nature of the photon pair generation process
implies that the period associated with the interference curve is
$\pi$.  Note also that for a sufficient departure from monochromatic
pumps in a realistic experimental situation, the joint spectral
amplitude may be computed by numerical integration of Eq.~\ref{eq:
JSA}. We have found that for pump bandwidths up to $\sim 0.7$nm,
Eq.~\ref{JSAinterf} may be used.

As an illustration, we will consider a specific example based on a
photonic crystal fiber with core radius $r=0.8658\mu$m and air
filling fraction $f=0.4$.   This fiber has a zero dispersion
frequency such that $\lambda_{zd}=2 \pi c/ \omega_{zd}=0.8089\mu$m.
Conditions i) through iii) from section~\ref{condiciones} are
fulfilled for the DP pump frequency with $\lambda_{DP}=2 \pi c/
\omega_{DP}=0.8089\mu$m as well as for one NDP frequency
pair with $\lambda_{NDP,1}=2 \pi c/\omega_{NDP,1}=0.7904\mu$m and
$\lambda_{NDP,2}=2 \pi c/\omega_{NDP,2}=0.8283\mu$m. We have
designed the source so that the three frequencies, which must be
mutually coherent, could be taken from the broad spectrum of a
femtosecond-duration pulse train, e.g. from a Ti:sapphire
oscillator.  Fig.~\ref{interf}(a) shows the singles spectrum,
computed numerically from Eq.~\ref{eq: JSA} assuming a pump spectral
amplitude of the form shown in Eq.~\ref{pumprect}, for a number of
different phase values $\theta$. Here we have assumed a peak pump
power of $P=5$W and likewise we have assumed that the three pump
spectral bands have the same spectral width (equal to $\Delta \lambda=0.5$nm) and where the amplitude of the central
one is higher by a factor of $\sqrt{2}$, compared to the other two. In
Fig.~\ref{interf}(a), the DP and two NDP spectral components are
indicated by vertical dashed lines.  It is clear from the figure
that as $\theta$ is increased from $0$ to $\pi/2$, the height of the
spectrum is reduced until emission is suppressed at $\theta=\pi/2$.
Fig.~\ref{interf}(b) shows the total emitted flux (given as the
integrated spectral flux) as a function of $\theta$.  This plot
explicitly shows the expected sinusoidal oscillations in the emitted
flux.

\begin{figure}[t]
\begin{center}
\centering\includegraphics[width=6.5cm]{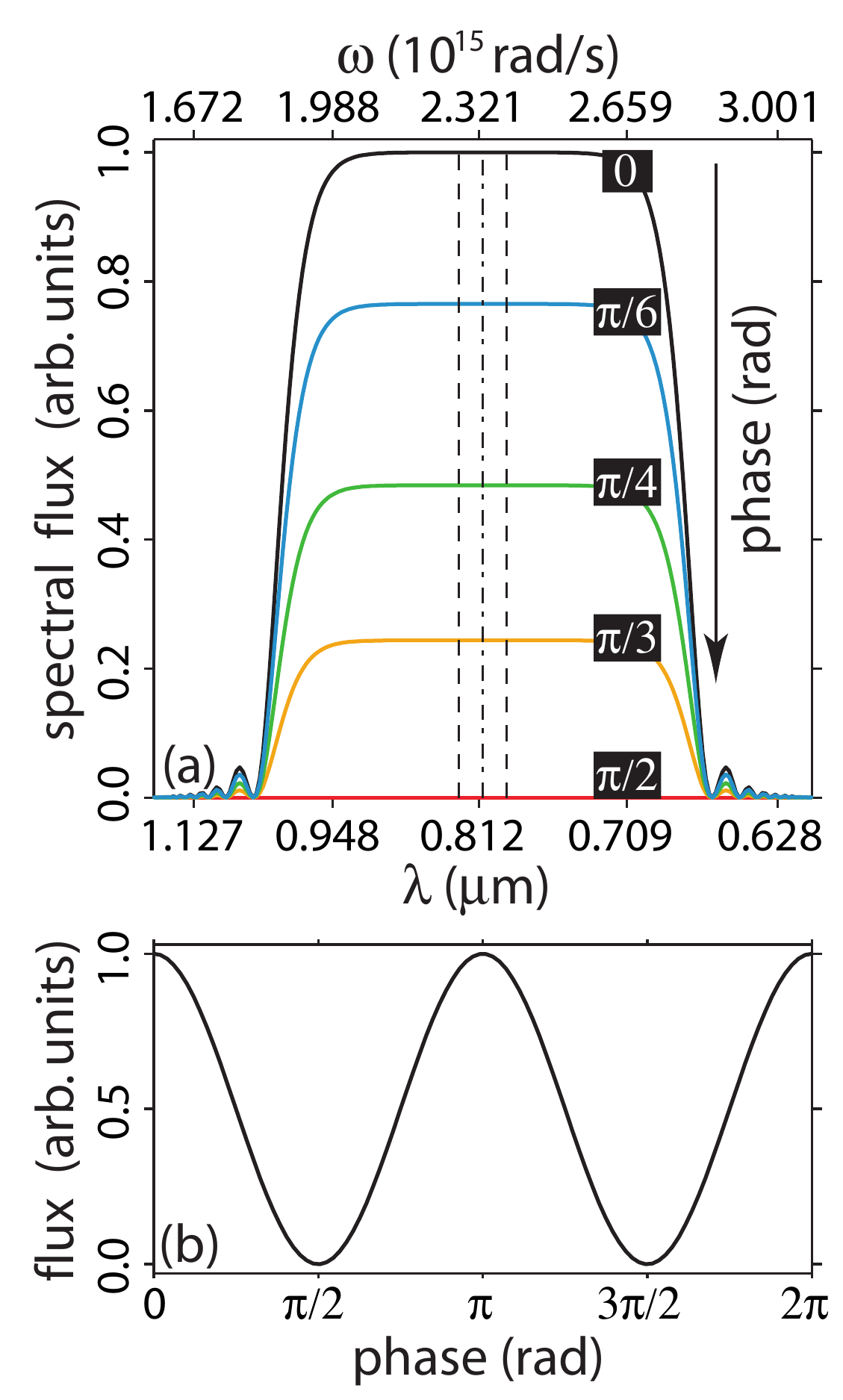}
\end{center}
\caption{(Color online) (a) Singles spectrum resulting for a fiber with
$r=0.8658\mu$m and $f=0.4$, pumped by three narrow spectral bands
centered at the frequencies corresponding to the DP and NDP regimes.
The curves shown were computed for different phase differences
(shown in black boxes) between the two pump configurations. (b)
Integrated spectra, proportional to the total emitted flux, as
a function of the phase difference between the two pump
configurations.}
\label{interf}       
\end{figure}

\section{Conclusions}

We have analyzed the spectral properties of photon pairs generated
by the process of spontaneous four wave mixing in single-mode fibers
and for narrow-band pumps.  Exploiting this analysis, we have
derived conditions under which ultra-broadband photon pair
generation with quasi-monochromatic pumps is possible. The resulting
two-photon states are highly entangled as quantified by the Schmidt
number and exhibit a particularly short correlation time.  We have
found that, for a given fiber, the attainable bandwidth is optimized
if perfect phasematching is attained for signal and idler
frequencies coinciding with the zero dispersion frequency. It is
possible to design a photon pair source that satisfies this
requirement which is based on degenerate, or on non-degenerate
pumps. These two regimes, referred to as DP and NDP in the paper,
lead to identical resulting two-photon states, revealing a
remarkable symmetry in the process of SFWM.   We have shown that
this symmetry leads to a quantum interference effect when the fiber is
pumped, for example, simultaneously by the pump frequencies
corresponding to the DP and NDP regimes. This symmetry also permits,
in the NDP regime, the generation of ultra-broadband photon pairs
without contamination due to spontaneous Raman scattering; note that
this represents a key concern in the design of fiber-based
photon-pair sources.  Although our theory can be applied to any fiber, we have focused our discussion on the use of
photonic crystal fibers, described through a step index, effective
medium dispersion model.   We have shown that for a given air
filling fraction in the cladding, the SFWM bandwidth is optimized
for core radii close to that which leads to the fulfilment of the
condition $k^{(4)}=0$.  Likewise, we have shown numerically that smaller air-filling fractions lead to greater SFWM
bandwidths; this is due to a weakening of waveguide dispersion
resulting from a lower nucleus-cladding index contrast.  We have
presented specific experimental designs, in some cases leading to
over $1000$nm of emitted SFWM bandwidth. We expect that these
results will be useful in the design of photon pair sources to be
used in the exploration of high-dimensional continuous-variable
entanglement.

\begin{acknowledgements}
KGP acknowledges support from Ph.D. dissertation scholarship (207537) from CONACYT-Mexico. ABU acknowledges support from
CONACYT-Mexico through grant 46370. RRR acknowledges CONACYT-Mexico for support through grant 46492.
\end{acknowledgements}


\end{document}